\documentclass[trackchanges, twocolumn]{aastex701}
\usepackage[flushleft]{threeparttable}

\usepackage{amsmath}	
\usepackage{caption}
\usepackage{subfig}

\begin{document}

\title{JWST Observations of Starbursts: The motion of dust - PAH kinematics of M82 and NGC 253 with JWST/MIRI spectroscopy}

\author[orcid=0000-0002-6460-3682,sname='Donnan']{Fergus R. Donnan}
\affiliation{Department of Astrophysics, University of California San Diego, 9500 Gilman Drive, San Diego, CA 92093, USA}
\email[show]{fdonnan@ucsd.edu} 

\author[orcid=0000-0002-4378-8534,sname='Sandstrom']{Karin Sandstrom}
\affiliation{Department of Astrophysics, University of California San Diego, 9500 Gilman Drive, San Diego, CA 92093, USA}
\email[]{kmsandstrom@ucsd.edu}  

\author[orcid=0000-0002-5480-5686,sname='Bolatto']{Alberto D. Bolatto}
\affiliation{Department of Astronomy, University of Maryland, College Park, MD 20742, USA}
\affiliation{Joint Space-Science Institute, University of Maryland, College Park, MD 20742, USA}
\email[]{bolatto@umd.edu}

\author[orcid=0000-0002-0846-936X,sname='Draine']{Bruce T. Draine}
\affiliation{Princeton University, Princeton, NJ 08540, USA}
\email[]{draine@astro.princeton.edu}

\author[orcid=0000-0003-0014-0508,sname='Duval']{Sara E. Duval}
\affiliation{Department of Physics \& Astronomy and Ritter Astrophysical Research Center, University of Toledo, Toledo, OH 43606, USA}
\email[]{saraduval018@gmail.com}  

\author[0000-0002-0560-3172]{Ralf S. Klessen}
\affiliation{Universität Heidelberg, Zentrum für Astronomie, Institut für Theoretische Astrophysik, Albert-Ueberle-Str. 2, D-69120 Heidelberg, Germany}
\affiliation{Universität Heidelberg, Interdisziplinäres Zentrum für Wissenschaftliches Rechnen, Im Neuenheimer Feld 205, D-69120 Heidelberg, Germany}
\email{klessen@uni-heidelberg.de}

\author[orcid=0000-0001-8782-1992,sname='Mills']{Elisabeth A.C. Mills}
\affiliation{Department of Physics and Astronomy, University of Kansas, 1251 Wescoe Hall Drive, Lawrence, KS 66045, USA}
\email[]{eacmills@ku.edu}  

\author[orcid=0000-0001-6325-9317,sname='Richie']{Helena M. Richie}
\affiliation{Department of Physics and Astronomy, University of Pittsburgh, 3941 O’Hara Street, Pittsburgh, PA 15260, USA}
\email[]{helenarichie@pitt.edu}  

\author[orcid=0000-0003-1545-5078,sname='Smith']{J.D.T. Smith}
\affiliation{Department of Physics \& Astronomy and Ritter Astrophysical Research Center, University of Toledo, Toledo, OH 43606, USA}
\email[]{jd.smith@utoledo.edu}  

\author[orcid=0000-0002-3158-6820,sname='Veilleux']{Sylvain Veilleux}
\affiliation{Department of Astronomy, University of Maryland, College Park, MD 20742, USA}
\affiliation{Joint Space-Science Institute, University of Maryland, College Park, MD 20742, USA}
\email[]{veilleux@umd.edu}

\author[0009-0001-6844-9758]{Patricia~A.~Arens}
\affiliation{Centre for Astrophysics and Supercomputing, Swinburne University of Technology, Hawthorn, VIC 3122, Australia}
\email{parens@swin.edu.au}

\author[0000-0002-3952-8588]{Leindert A. Boogaard} \affiliation{Leiden Observatory, Leiden University, PO Box 9513, NL-2300 RA Leiden, The Netherlands}
\email{}

\author[0000-0002-9511-1330, sname='Cronin']{Serena A. Cronin}
\affiliation{Department of Astronomy, University of Maryland, College Park, MD 20742, USA}
\email{cronin@umd.edu}

\author[0000-0002-5782-9093]{Daniel~A.~Dale}
\affiliation{Department of Physics and Astronomy, University of Wyoming, Laramie, WY 82071, USA}
\email{ddale@uwyo.edu}

\author[0000-0001-6527-6954, sname='Emig']{Kimberly L. Emig}
\affiliation{National Radio Astronomy Observatory, 520 Edgemont Road, Charlottesville, VA 22903, USA}
\email{kemig@nrao.edu}

\author[0000-0002-2775-0595, sname='Herrera-Camus']{Rodrigo Herrera-Camus}
\affiliation{Departamento de Astronomía, Universidad de Concepción, Barrio Universitario, Concepción, Chile}
\affiliation{Millennium Nucleus for Galaxies (MINGAL), Concepción, Chile}
\affiliation{Max-Planck-Institut für extraterrestrische Physik, Giessenbachstrasse 1, 85748 Garching, Germany}
\email{rodrherrera@udec.cl}

\author[0000-0001-8490-6632, sname='Lai']{Thomas S.-Y. Lai}
\affiliation{IPAC, California Institute of Technology, 1200 East California Boulevard, Pasadena, CA 91125, USA}
\email{shaoyu@ipac.caltech.edu}

\author[0000-0002-2644-0077]{Sebastian Lopez} 
\affiliation{School of Earth and Space Exploration, Arizona State University, P.O. Box 876004, Tempe, AZ 85287, USA}
\email{lopez.764@osu.edu}

\author[0000-0003-4023-8657, sname='Lenkić']{Laura Lenkić}
\affiliation{IPAC, California Institute of Technology, 1200 East California Boulevard, Pasadena, CA 91125, USA}
\email{llenkic@ipac.caltech.edu}

\author[orcid=0000-0003-2508-2586,sname='Levy']{Rebecca C. Levy}
\affiliation{Space Telescope Science Institute, 3700 San Martin Drive, Baltimore, MD 21218, USA}
\email[]{rlevy.astro@gmail.com}  

\author[orcid=0000-0002-4677-0516,sname='Mayya']{Divakara Mayya}
\affiliation{Instituto Nacional de Astrof{\'\i}sica, \'Optica y
Electr\'onica Luis Enrique Erro 1, Tonantzintla, 72840, Puebla,
Mexico}
\email[]{ydm@inaoep.mx}

\author[0000-0001-9436-9471, sname='Meier']{David S. Meier}
\affiliation{New Mexico Institute of Mining and Technology, 801 Leroy Place, Socorro, NM 87801, USA}
\email{David.Meier@nmt.edu}

\author[0009-0005-8382-0614]{Kaitlyn E. Sheriff}
\affiliation{Department of Physics and Astronomy, University of Kansas, 1251 Wescoe Hall Drive, Lawrence, KS 66045, USA}
\email[]{kaitlyn.sheriff@ku.edu}

\author[orcid=0000-0003-3633-0098,sname='Siwakoti']{Utsav Siwakoti}
\affiliation{Department of Physics and Astronomy, University of Kansas, 1251 Wescoe Hall Drive, Lawrence, KS 66045, USA}
\email{siwakotiutsav@ku.edu}

\author[0000-0003-0605-8732, sname='Skillman']{Evan Skillman}
\affiliation{Minnesota Institute for Astrophysics, University of Minnesota, 116 Church Street SE, Minneapolis, MN 55455, USA }
\email{skill001@umn.edu}

\author[orcid=0000-0003-1356-1096,sname='Tarantino']{Elizabeth Tarantino}
\affiliation{Space Telescope Science Institute, 3700 San Martin Drive, Baltimore, MD 21218, USA}
\email{etarantino@stsci.edu}

\author[orcid=0000-0003-4209-1599,sname='Teng']{Yu-Hsuan Teng}
\affiliation{Department of Astronomy, University of Maryland, College Park, MD 20742, USA}
\email{yhteng@umd.edu}

\author[orcid=0000-0002-5877-379X,sname='Villanueva']{Vicente Villanueva}
\affiliation{Instituto de Estudios Astrofísicos, Facultad de Ingeniería y Ciencias, Universidad Diego Portales, Av. Ejército Libertador 441, 8370191 Santiago, Chile}
\affiliation{Millennium Nucleus for Galaxies (MINGAL), Concepción, Chile}
\email{vicente.avl365@gmail.com}

\author[orcid=0000-0003-4793-7880,sname='Walter']{Fabian Walter} 
\affiliation{Max Planck Institut für Astronomie, Königstuhl 17, D-69117 Heidelberg, Germany}
\email{walter@mpia.de}

\author[orcid=0000-0001-5434-5942,sname='Van der Werf']{Paul P.~van der Werf} 
\affiliation{Leiden Observatory, Leiden University, PO Box 9513, 2300 RA Leiden, The Netherlands}
\email{pvdwerf@strw.leidenuniv.nl}







\begin{abstract}
We present an analysis of dust kinematics in the local starburst galaxies M82 (NGC 3034) and NGC 253 using Polycyclic Aromatic Hydrocarbon (PAH) features observed with JWST/MIRI spectroscopy. We are able to produce high-quality velocity maps of the 5.2\,$\mu$m, 6.2\,$\mu$m, and 11.3\,$\mu$m PAH features, as well as numerous lines of molecular gas via H$_2$ rotational transitions and ionized gas from [\ion{Ne}{2}] and H recombination lines. Given the field of view and inclination, we trace a rotating disk in M82 where we observe a steep rise in the velocities followed by flattening, typical of galaxy rotation curves. In NGC 253, however, the 6.2~$\mu$m and 11.3~$\mu$m PAH features trace the launching region of the outflow, firmly within the starburst region. We find the ionized gas also shows outflow contributions in NGC 253 while the warm molecular gas is dominated by rotation. Additionally, the molecular gas outflow velocities are lower than the ionized gas, with the 11.3~$\mu$m PAH feature consistent with the ionized gas rather than the molecular gas. We therefore suggest that PAHs are more closely associated with the ionized gas rather than the molecular at the base of the galaxy outflow, where larger scale imaging shows comparable morphology between PAHs and HI. The 6.2~$\mu$m PAH feature has an even higher outflow velocity for both galaxies, possibly suggesting preferential ionization of the PAHs within the faster-moving hot phase of the base of the outflow.

\end{abstract}



\section{Introduction}

Dust is a fundamental component of the interstellar medium (ISM) of galaxies, regulating the thermal balance and chemistry, and obscuring much of the star-formation activity \citep[e.g.][]{Draine2003, Tielens2005, Zavala2021}. Unlike the gas phase which emits through various emission lines, measuring the kinematics of dust is highly challenging. However, it is now possible with JWST spectroscopy of Polycyclic Aromatic Hydrocarbons (PAHs), which are the smallest dust grains in the ISM, contributing up to $5\%$ of the dust mass \citep[e.g.][]{Draine2007} and up to $20\%$ of the infrared luminosity \citep[][]{Smith2007}. After excitation of these molecules from UV photons, they give rise to a multitude of broad emission features in the mid-infrared \citep[e.g.][]{Peeters2002, Tielens2008, Li2020}. PAH features are, however, intrinsically broad, with widths orders of magnitude larger than any velocity shift. Moreover, the shape of the intrinsic profile is not well defined and can vary in different astrophysical environments \citep[e.g.][]{Peeters2002, VanDePutte2025}.

To overcome these challenges, \citealt{Donnan24b} used Principal Component Analysis (PCA) tomography to derive the first velocity maps of PAH features in galaxies. PCA is a data-driven technique that reorders data cubes into linear principal components, enabling one to measure velocity shifts much smaller than the width of the feature without the need of modeling the intrinsic profile of the emission feature. More recent work has presented PAH kinematics in AGN, where kinematic evidence for dust in the outflow of AGN has been found \citep[][]{Donnan2026c, Donnan2026}. 

Perhaps the best extragalactic laboratories in which to measure the motion of PAHs are local starburst galaxies. Starbursts exhibit high star-formation efficiencies \citep[e.g.][]{Leroy2013, Kennicutt2021, Fisher2022}; the feedback from which drives large scale galactic outflows \citep[see][for a recent review]{Thompson2024} with PAH emission detected out to 2 kpc in M82 \citep[][]{Engelbracht2006, Beirao2015, Lopez2025, Cronin2026}; and even further in higher-$z$ starbursts \citep[$\sim35$ kpc;][]{Veilleux2025}.
Galactic outflows are multi-phase and play a key role in the baryon cycle, injecting high metallicity material into the intergalactic medium (IGM) and circumgalactic medium (CGM) \citep[e.g.,][]{Veilleux2005, Strickland2009, Leroy2015, Veilleux2020} and potentially quenching star formation in the galaxy through the ejection of molecular gas \citep[e.g.,][]{Chisholm2016}. These outflows may be driven by radiation pressure on dust grains \citep[][]{Murray2005, Thompson2005}, making measuring the kinematics of dust extremely valuable to understanding starburst driven outflows.

The velocity information of PAHs provides a unique constraint on their properties where differences in the motion of various PAH bands trace distinct phases. For example, \citealt{Donnan2026} found that the 3.3\,$\mu$m PAH feature was not kinematically present in AGN outflows but the 11.3\,$\mu$m PAH feature was present, independently confirming a lack of small PAHs in AGN outflows suggested from the flux ratios \citep[][]{Rigopoulou2024, Garcia-Bernete2024b, Garcia-Bernete2026}. 

\citealt{Bolatto2024} showed that the morphology of the PAH emission in the outflow of M82 more closely matched that of the ionized gas rather than the molecular gas (see also \citealt{Cronin2026}), suggesting that PAHs may be coupled with the ionized gas. However, \citealt{Lopez2025} and \citealt{Villanueva2025} reported the typical CO-PAH relation held across the outflow, suggesting the PAHs are coupled with the cool molecular gas. PAH kinematics will provide a crucial discriminant on how the PAHs are coupled to the gas phases.

With the spatial resolution of JWST/MIRI MRS, we are able to probe the kinematics of the launching region of the outflow ($\sim30$pc) within the starburst, in the ionized, molecular and now dust phase, providing new constraints on how these large scale ($\sim$kpc) starburst driven outflows are launched.

In this paper we present PAH kinematic maps of M82 and NGC 253. We first summarize the technique of PCA tomography in Section \ref{sec:PCA} before presenting velocity maps in Section \ref{sec:Results}. Finally, we discuss the implications of our results in Section \ref{sec:Discuss}.

\section{Observations}

Observations of M82 and NGC 253 were obtained through GO Program 1701 (P.I. A. Bolatto), and the data were obtained from the Mikulski Archive for Space Telescopes at the DOI:\href{https://archive.stsci.edu/doi/resolve/resolve.html?doi=10.17909/ta3c-0426}{10.17909/ta3c-0426}. The galaxies were observed with NIRCam and MIRI imaging as well as MIRI MRS spectroscopy. In this paper we focus on the MIRI MRS spectroscopy \citep{Duval2026}.

\begin{figure*}
	\includegraphics[width=\textwidth]{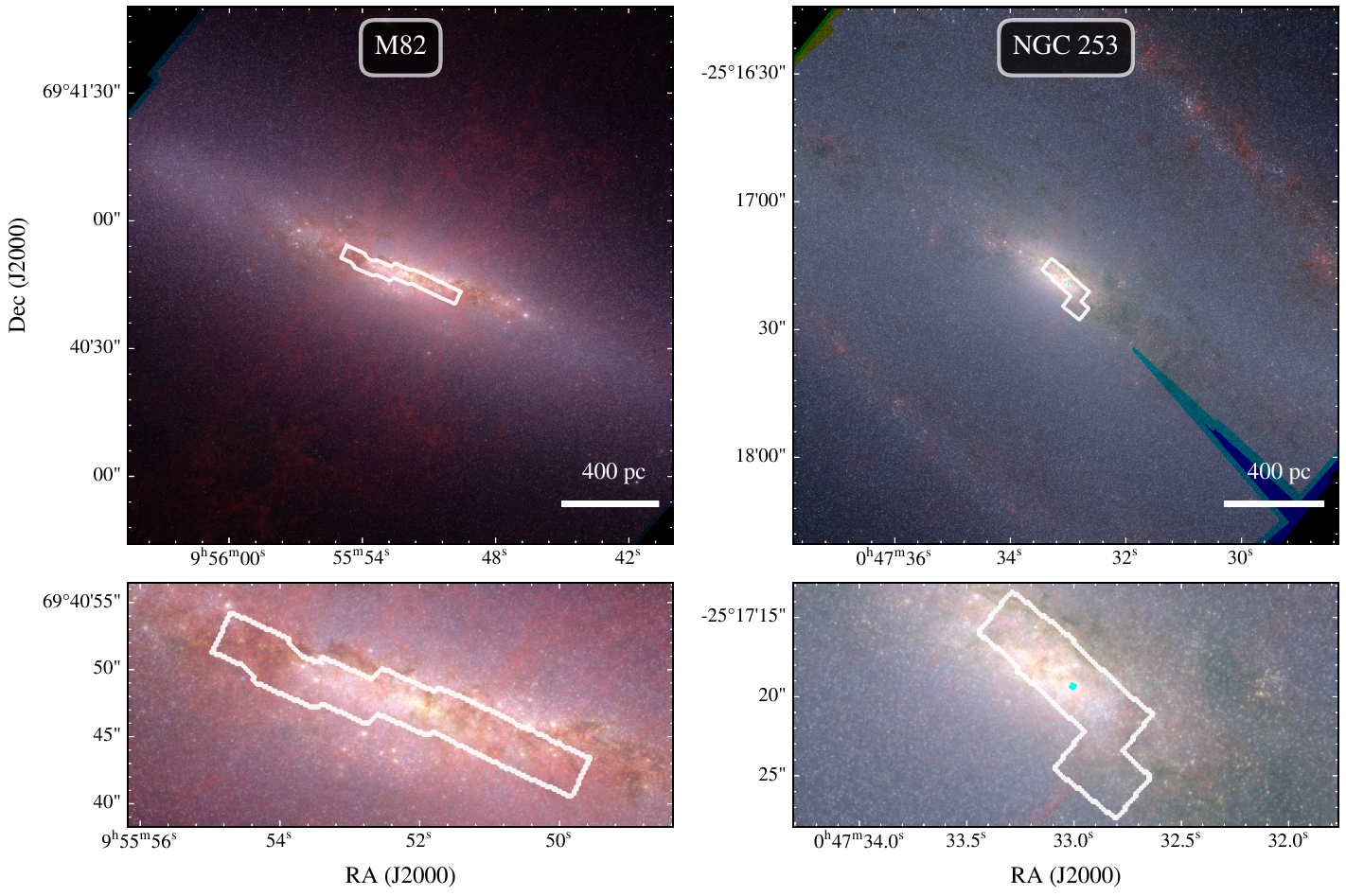}
    \caption{Three-color NIRCAM images of M82 and NGC 253 using the F140M, F212N and F335M filters for blue, green and red respectively. The field of view of Channel 3 of the MIRI MRS observations are shown with the white box. The lower panels show a zoom of the central regions. Note that there is some saturation of the images for NGC 253 from the cluster region that we mask as discussed in Section \ref{sec:IntProfile}.}
    \label{fig:Images} 
\end{figure*}

We show three-color images of M82 and NGC 253 in Fig.~\ref{fig:Images}, where we use the F140M, F212N and F335M NIRCAM filters \citep[][]{Bolatto2024, Cronin2026, Levy2026} for blue, green and red respectively. For more details about the imaging see \citealt{Cronin2026}. In these images the F335M filter is dominated by the 3.3\,$\mu$m PAH emission which is prominent in the galaxy outflow while the F212N is centered on the 2.12\,$\mu$m H$_2$ line. The field of view of the MIRI MRS mosaics are shown with the white box. 

The MIRI MRS data were reduced using pipeline version 2.0.0 with CDRS version 13.0.6. For full details of the data reduction process see \citealt{Duval2026}. 

\section{PCA Tomography}
\label{sec:PCA}
Principal Component Analysis (PCA) tomography is a technique to decompose astronomy data cubes (2 spatial dimensions and 1 spectral dimension) into linear principal components \citep{Heyer1997, Steiner2009}. PCA is a dimensionality reduction technique that identifies a series of orthonormal spectra (eigenspectra), that show some specific spectral shape with spatial correlation. In the case of emission features that are subject to velocity shifts, the first principal component contains the rest frame emission while typically the second identifies regions where the feature is blueshifted or redshifted and therefore allows one to construct velocity maps.

The advantage of PCA tomography over more traditional techniques such as modeling each spaxel is that it is a data-driven technique and one that can identify the signature of kinematics without prior knowledge of the intrinsic emission feature. Moreover, PCA maximizes the information in the data cube held in the spatial axes to detect Doppler shifts. This is particularly useful for PAH features, where the intrinsic profile is not well defined \citep[e.g.][]{Peeters2002, Pasquini2023, VanDePutte2025}. Moreover, any velocity shift on the order of $\sim 100$\,km\,s$^{-1}$ is significantly lower than the width of a PAH feature. Using PCA tomography, the first velocity maps of PAH features were presented by \citealt{Donnan24b}.

We first produce data sub-cubes of individual emission features. Before the PCA decomposition is applied to the data cube of a given emission feature, a local continuum is first subtracted around the feature. We do this by fitting a linear function in $f_{\nu}$ to the 15 wavelength channels of either side of the sub-cube. As presented in \citealt{Donnan2026c}, the exact continuum used does not make any significant difference to the PCA decomposition and the subsequent inferred velocity map (see their Fig.~B.1.). This isolates the emission feature, preventing the PCA from detecting differences in continuum rather than kinematics. We then mask any emission lines that appear on top of the PAH features. The mean spectrum is then also subtracted from each spaxel as a standard practice for PCA analysis. The spatial axes of the cube, $(x,y)$, are then collapsed into a single coordinate, $\beta$, where
\begin{equation}
\label{eqn:CoordTrans}
    \beta = N_x (x-1) +y,
\end{equation}
where $N_x$ is the total number of spatial pixels along the x direction. The PCA decomposition can be written as 
\begin{equation}
\label{eqn:PCA}
    \mathbf{T}_{\beta, k} = \mathbf{D}_{\beta, \lambda} \cdot \mathbf{E}_{\lambda, k}, 
\end{equation}
where $\mathbf{D}_{\beta, \lambda}$ is the original data cube with the collapsed spatial axes. In this transformation, $\mathbf{T}_{\beta, k}$ are the tomograms of each principal component, $k$, which describes the spatial correlation of a given spectral feature, $\mathbf{E}_{\lambda, k}$, known as the eigenspectra.

To construct a velocity map, we assume that the eigenspectrum of the first principal component is representative of the rest frame emission while the higher order components contain information on velocities with respect to the rest frame. We can therefore measure a velocity shift by reconstructing the data cube with the first four principal components and measuring the wavelength shift, $\Delta \lambda_{x,y}$, of each spaxel to the first principal component. The peak of the first eigenspectrum and the wavelength where the second eigenspectrum crosses zero corresponds to the central wavelength, $\lambda_0$, which we use to infer the velocity of each spaxel as 
\begin{equation}
    V_{x,y} = c \frac{\Delta \lambda_{x,y}}{\lambda_0},
\end{equation}
where $c$ is the speed of light. We verified this by testing against a more traditional method of fitting Gaussian profiles in Appendix \ref{sec:LineFits}.

To also produce a velocity error for each spaxel we repeat the PCA decomposition $N$ times, where each time we create a new set of data by resampling the data from a normal distribution, using the original data values as the mean and the data errors as the standard deviation. By calculating the standard deviation of the $N$ velocities for each spaxel, we measure a velocity error due to the noise in the spectrum.

\subsection{Intrinsic profile vs Velocity Shifts}
\label{sec:IntProfile}
PCA tomography is a data-driven technique that is able to detect variations in the emission profile with spatial correlation. In the case of pure Doppler shifts due to the velocity of the emitter, the velocity typically presents in the second principal component. However, with PAH emission, the intrinsic profile of the feature can vary between different physical environments \citep[e.g.][]{Peeters2002, Diedenhoven2004, Pasquini2023, VanDePutte2025}. Therefore it may be unclear whether variations seen are due to velocity shifts or variations in the intrinsic profile, however, we can look at the shape of the eigenspectra to see if they behave as expected, where the profile shifting is purely due to velocity.

The spectrum of a PAH feature, $F_{\rm rest}\left(\lambda'\right)$, shifted due to velocity, $v$, can be expressed as 
\begin{equation}
    F_{\rm obs}\left(\lambda\right) = F_{\rm rest}\left(\lambda\left(1+\frac{v}{c}\right)\right),
\end{equation}
where $c$ is the speed of light and $\lambda' = \lambda\left(1+\frac{v}{c}\right)$. As $v\ll c$, we can do a first order Taylor expansion and write 
\begin{equation}
    F_{\rm obs}\left(\lambda\right) \approx F_{\rm rest}\left(\lambda\right) + \frac{v}{c}\frac{dF_{\rm rest}}{d\lambda} \lambda .
\end{equation}
Therefore a Doppler shift can be approximated as a linear function where the first term is the rest frame emission and the second is proportional to the derivative of the first term. As PCA is a linear decomposition, we expect the eigenspectrum of the second principal component to look like the derivative of the first component. We can write the flux at a given spaxel, $i$, as a sum of the principal components
\begin{equation}
\label{eqn:PCA_ex}
    F_i\left(\lambda\right) = \bar{F}\left(\lambda\right) + a_i^{(1)}\phi_1(\lambda) + a_i^{(2)}\phi_2(\lambda) + ..
\end{equation}
where $\bar{F}\left(\lambda\right)$ is the mean spectrum which is subtracted from each spaxel in the data cube as a standard practice before applying the PCA decomposition, $a_i^{(1)}$, is the value of the tomogram of spaxel $i$ for the first principal component while $\phi_1(\lambda)$ is the eigenspectrum of the first component. The second principal component consists of $a_i^{(2)}$, for spaxel $i$ and $\phi_2(\lambda)$ is the second eigenspectrum. Comparing equation \ref{eqn:PCA_ex} to the Taylor expansion due to a Doppler shift, if the kinematics are responsible for the profile variations, we would expect the eigenspectrum of the second principal component to be proportional to the derivative of the first
\begin{equation}
\phi_2(\lambda) \propto \frac{dF_{\rm rest}}{d\lambda}\lambda \propto \frac{dF_{\rm rest}}{d\lambda}
\end{equation}
as $\lambda \approx\lambda_0$ and therefore is approximately constant over the width of a spectral feature, and the tomogram represents the velocity map, i.e.,  $a_i^{(2)}\propto v_i$. 

We show an example where there are no velocity shifts but rather the intrinsic profile varies spatially in Appendix \ref{sec:IntProfile2}, using the MIRI MRS data of the Orion bar \citep[][]{Berne2022}.

For our data of M82 and NGC 253, we find that the eigenspectrum of the second principal component of the 5.2~$\mu$m, 6.2\,$\mu$m, 11.3\,$\mu$m and 17\,$\mu$m PAH features match that of the derivative of the first component suggesting Doppler shifts dominate over any possible changes in the intrinsic profile. We show this in Appendix \ref{sec:IntProfile2}. For NGC 253, if we do not mask the cluster region (Region 5 from \citealt{Leroy2018}), we find that the second principal component differs from the derivative of the first, particularly for the 6.2\,$\mu$m PAH suggesting PAH reprocessing in this region. For the purposes of constructing velocity maps, we mask the cluster region in this work.

We also tested the PCA method against fitting a Gaussian for the emission lines, to see if the inferred velocity maps are the same. We do indeed find that they are identical as presented in Appendix \ref{sec:LineFits}.

\section{Results}
\label{sec:Results}
\subsection{Velocity Maps}

Considering the high signal to noise of the data, we are able to produce velocity maps for many of the PAH features beyond just the brightest, namely the 5.2\,$\mu$m, 6.2~$\mu$m, 11.3\,$\mu$m and 17\,$\mu$m features. For the 17\,$\mu$m feature, we use the 16.4\,$\mu$m sub-feature to measure kinematics as the full complex is much too broad. 
Additionally, we ran the PCA analysis for the 7.7\,$\mu$m, 8.6\,$\mu$m, 11.0\,$\mu$m features. In the case of the 7.7\,$\mu$m PAH, the feature was too broad to infer any kinematics while the 8.6\,$\mu$m and 11.0\,$\mu$m features showed a second eigenspectrum not consistent with the derivative of the first eigenspectrum suggesting some variations in the intrinsic profile. 

\begin{figure*}
	\includegraphics[width=\textwidth]{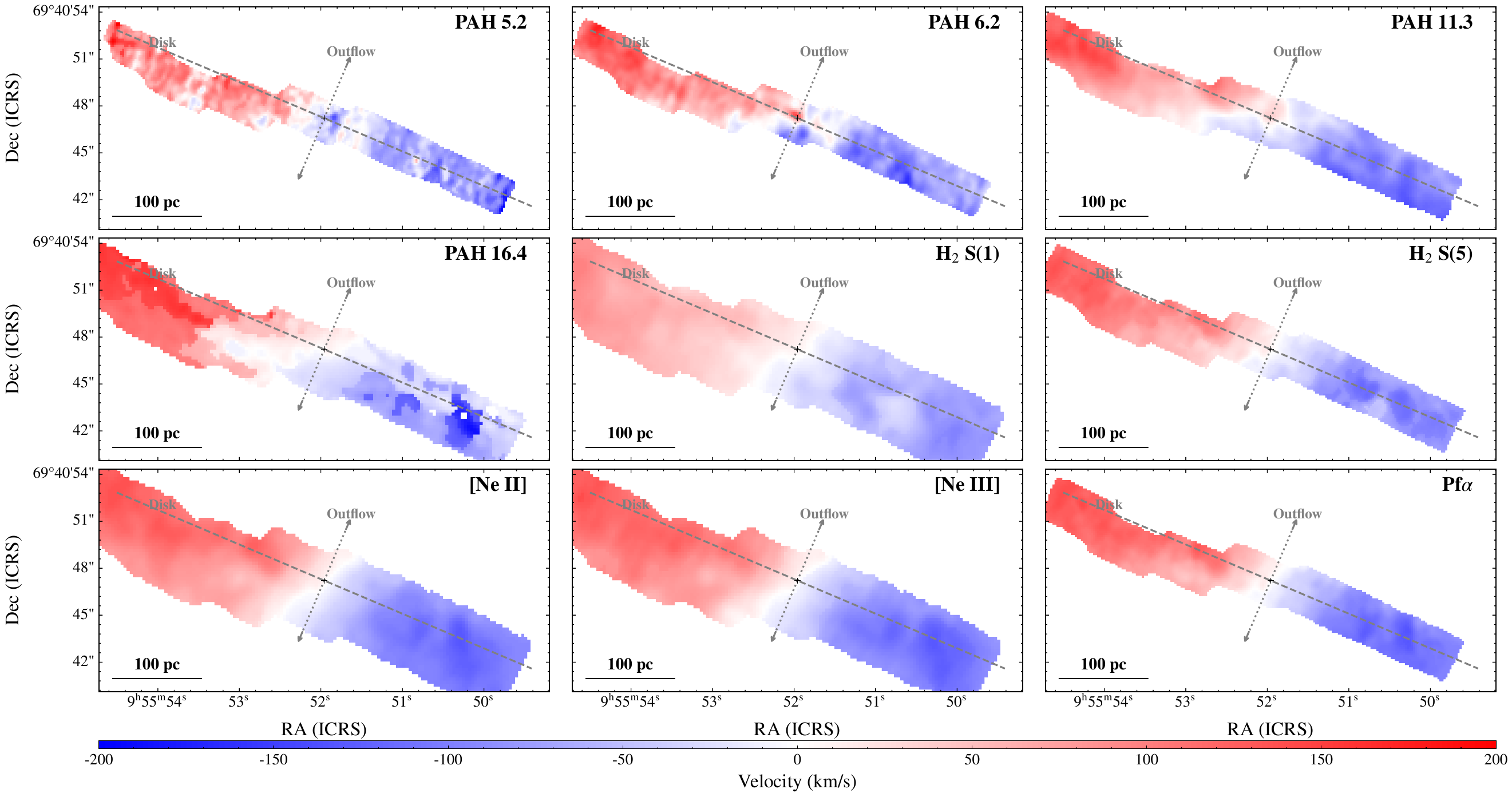}
    \caption{Velocity maps of various emission features in M82. We show maps of various PAH features, rotational transitions of H$_2$, and ionized gas via [\ion{Ne}{2}], [\ion{Ne}{3}] and the hydrogen recombination line Pf$\alpha$. The position angle of the major axis of the galaxy disk is shown with the gray dashed lines, while the outflow direction, which is perpendicular to the disk, is shown in the dotted line. We mask spaxels with a velocity error $>100$~km~s$^{-1}$. Note the presence of the outflow in the 6.2\,$\mu$m and 11.3\,$\mu$m PAH velocity maps. }
    \label{fig:VelMapsM82} 
\end{figure*}

\begin{figure*}
	\includegraphics[width=\textwidth]{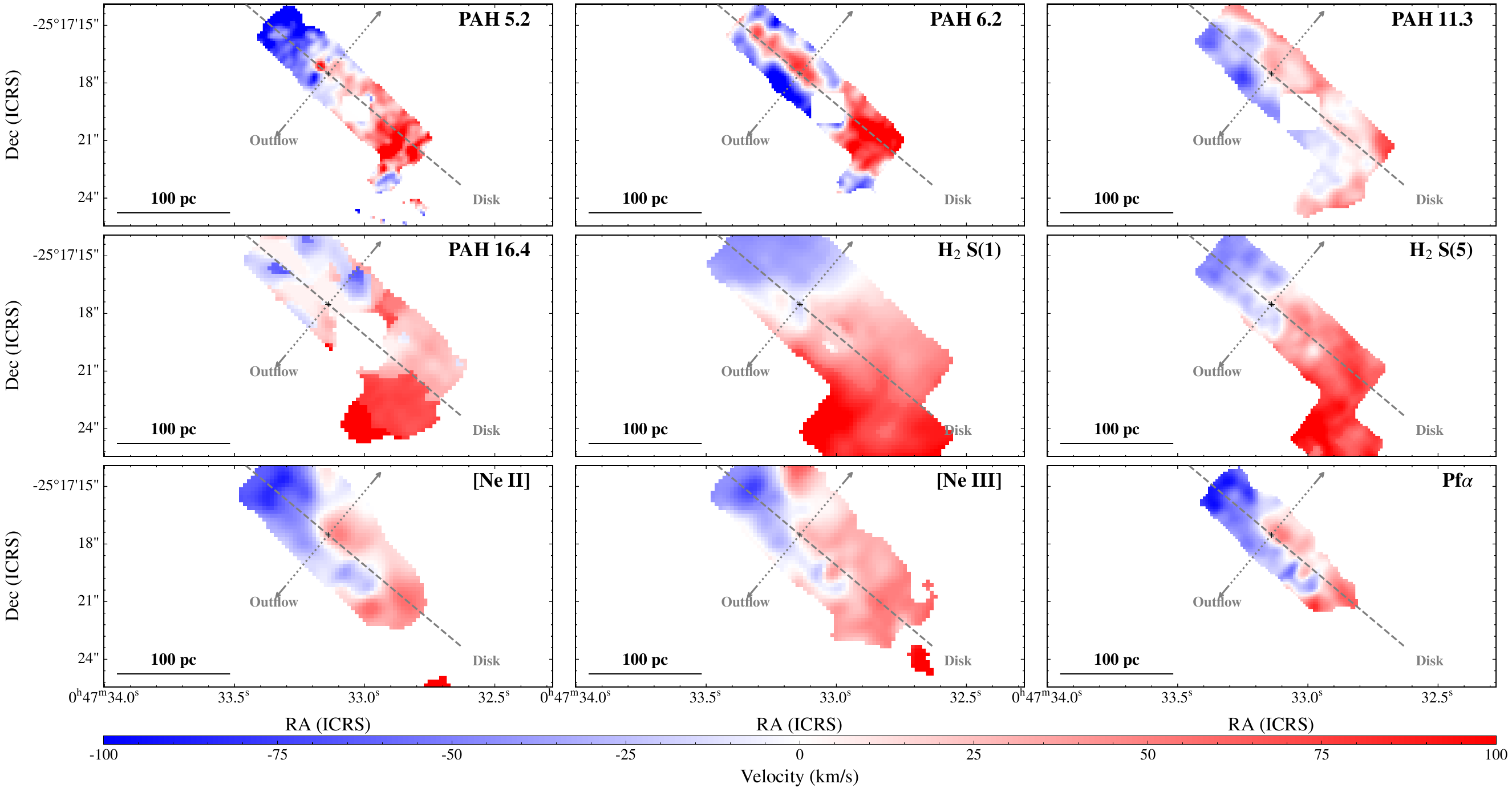}
    \caption{The same as Fig.~\ref{fig:VelMapsM82} but for NGC 253. Here we see velocities of the 6.2\,$\mu$m and 11.3\,$\mu$m PAH features along the outflow axis compared to the other lines which have a stronger disk contribution to the velocity maps. }
    \label{fig:VelMapsNGC253} 
\end{figure*}

The velocity maps are shown in Fig.~\ref{fig:VelMapsM82} and Fig.~\ref{fig:VelMapsNGC253} for M82 and NGC 253, respectively. Note that the spatial resolution of MIRI ranges from $\sim0.13''-0.2''$ per spaxel which corresponds to a spatial scale of $\sim2.2-3.4$pc.
We also plot the kinematics of the H$_2$ S(1) and S(5) rotational transitions (we also measure the kinematics of the S(3)) line to probe the molecular gas as well as [\ion{Ne}{2}], [\ion{Ne}{3}] and Pf$\alpha$ to probe the ionized gas. We have displayed the position angles of the disk and outflow for M82 and NGC 253, constrained from CO kinematics \citep[][]{Leroy2015} and stellar kinematics \citep[][]{Muller-Sanchez2010}.

We find that the PAH, H$_2$ and ionized gas kinematics of M82 strongly trace rotation for all the features, where the outflow appears weak in our velocity maps. This is likely partly due to the high inclination \citep[$\sim$80$^{\circ}$,][]{Leroy2015} of the disk, but also the limited FOV along the outflow direction. There are, however, some differences in the inner regions of the galaxy, where the position angle of the PAH velocity map, particularly the 6.2\,$\mu$m and 11.3\,$\mu$m PAHs, appears to show redshifted emission to the north and blueshifted emission to the south of the nucleus. This is spatially coincident with the base of the large scale galactic outflow, with the velocity and orientation matching what is observed for the outflow. It is worth noting that given the limited FOV along the outflow axis, we are only probing $\sim30$pc from the disk, while the outflow extends to many kpc, meaning we are firmly within the starburst region rather than the extended outflow. We investigate this further in Section \ref{sec:Profiles}.

The kinematics of NGC 253 are more varied, with the rotation of the H$_2$ S(1) and S(5) lines matching that of rotation in the galaxy disk; however, the other emission features show clear differences. Most notably, the 6.2~$\mu$m and 11.3\,$\mu$m PAH features instead have a position angle perpendicular to the disk and consistent with the outflow. 

The southwest of the field of view of our MRS observations coincide with the streamer as discussed in \citealt{Walter2017, Cronin2025}. Unfortunately this region is largely masked in our velocity maps due to the low signal-to-noise and thus high velocity errors. However there is a hint of blueshifted velocities in the 6.2\,$\mu$m PAH velocity map while the H$_2$ lines appear to show the disk rotation in the SW region.


\subsection{Disk and Outflow Kinematics}
\label{sec:Profiles}
\begin{figure*}
	\includegraphics[width=\textwidth]{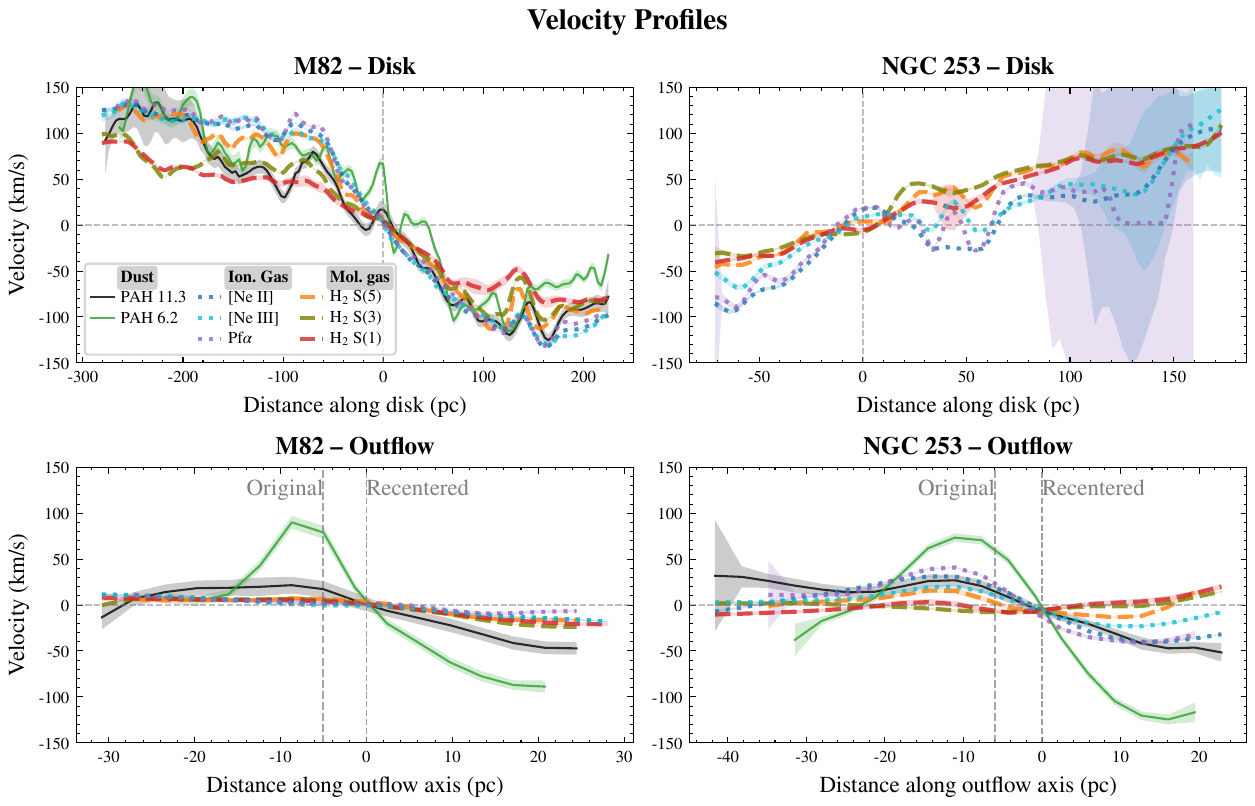}
    \caption{1D velocity profiles of the disk and outflow axes. Top panels show the velocity profiles extracted through 2$''$ wide apertures along the major axis of the disk for M82 and NGC 253 with various emission features. The bottom panels show the velocity profiles through 1$''$ wide apertures of the outflow, taken perpendicular to the major axis of the disk. The PAH features are shown with a solid line, the H$_2$ lines are shown with a dashed line while the ionized gas tracers are shown with a dotted line. The shaded regions show the 1$\sigma$ errors in the velocity. We omit the 5.2\,$\mu$m and 16.4\,$\mu$m PAH features from this plot as there may be non-negligible contributions to the inferred velocity from variations in the intrinsic profile as demonstrated in Fig.~\ref{fig:VelDiff}.  We also omit the 6.2\,$\mu$m and 11.3\,$\mu$m PAH features for the disk profiles of NGC 253 as these features are dominated by the outflow.
    It is worth noting that these are only statistical errors from the Monte Carlo repeats of the PCA process and do not account for calibration uncertainties and therefore should be taken as lower limits of the errors. In the bottom panels the vertical lines show the position of the nucleus labeled ``Original'' and the adjusted center position labeled ``Recentered'' to coincide with where the sign of the velocity flips.}
    \label{fig:VelProfiles} 
\end{figure*}

We extract 1D velocity profiles of the galaxy disk by measuring the average velocity of an aperture with a width of 2$''$ ($\sim$35 pc) along the major axis of the disk. These are shown in the top panels of Fig.~\ref{fig:VelProfiles}. We find that the velocity profiles of M82 behave as expected, showing a steep increase in the inner regions before leveling off at $\sim$ 100 pc. The velocity of the PAH features matches the ionized gas well as traced by [\ion{Ne}{2}] and Pf$\alpha$ (HI 6-5) as well as the H$_2$ S(5) line. However, the lower J rotational transitions of H$_2$, namely the S(1), reaches a lower velocity. We discuss possible reasons for this in Section \ref{sec:Discuss}. 

The disk velocity profiles of NGC 253 are less coherent, where the 6.2\,$\mu$m and 11.3\,$\mu$m PAHs are dominated by the outflow and so show no clear sign of rotation. The warm molecular and ionized gas are dominated by rotation with no clear discrepancies between the H$_2$ rotational transitions like in M82.

NGC 253 shows clearer evidence of non-circular motion, matching the direction of the galactic outflow \citep[][]{Nico2019, Cronin2025} in the velocity maps (Fig.~\ref{fig:VelMapsNGC253}) and the 1D profile extracted perpendicular to the disk (Fig.~\ref{fig:VelProfiles}) than M82. This suggests we are observing the launching of the large scale galaxy outflow at its base. We find that the 6.2\,$\mu$m PAH shows the highest line of sight velocity, reaching $\sim75$\,km\,s$^{-1}$  while the 11.3\,$\mu$m PAH, consistent with the ionized gas as traced by [\ion{Ne}{2}] and Pf$\alpha$, reaches a lower velocity of $\lesssim25$\,km\,s$^{-1}$. It is worth noting that although $25$\,km\,s$^{-1}$ is lower than the $V_{\rm max}\sim40$\,km\,s$^{-1}$ from our testing in Appendix \ref{sec:Mock}, we do detect spaxels with velocities lower than $V_{\rm max}$ as this is the maximum velocity not the limit.

\begin{figure}
	\includegraphics[width=\columnwidth]{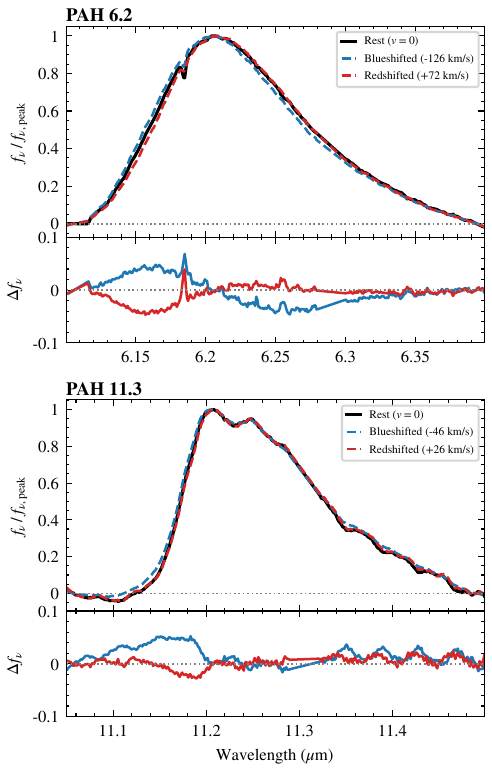}
    \caption{Profiles of the 6.2\,$\mu$m and 11.3\,$\mu$m PAH features at the peak velocities (red and blue) along the outflow axis of NGC 253 (see Fig.~\ref{fig:VelProfiles}) compared to the rest frame profile. The residual panels show the difference between each profile and the rest frame profile. Note that the profiles have been normalized to their peaks for visualization. The absorption feature at $\sim5.18\mu$m is likely the 110-101 ro-vibrational transition of H$_2$O, which is particularity strong in the nucleus giving rise to the spikes in the $\Delta f_{\nu}$ panel.
    This figure demonstrates how subtle the shifts in the profile are compared to its width and thus techniques such as PCA tomography are necessary to extract the velocity information from the data cube.}
    \label{fig:Prof} 
\end{figure}

The launching of the outflow is much less clear in the molecular gas; however, the H$_2$ S(5) line shows some evidence, reaching $\sim 20$\,km\,s$^{-1}$. The increased presence of the outflow in the higher-J transitions of H$_2$ suggests the molecular gas that is launched may be at a higher temperature than the disk. The velocity begins to decrease for all the features after peaking at $\sim10$ pc from the mid-plane of the galaxy. For reference we show the profiles of the 6.2\,$\mu$m PAH and 11.3\,$\mu$m PAH in the outflow regions of NGC 253 in Fig.~\ref{fig:Prof}.

While the outflow is less clear in M82, the 6.2\,$\mu$m PAH and 11.3\,$\mu$m PAH show tentative kinematic evidence, with again the 6.2\,$\mu$m PAH reaching a higher velocity of $\gtrsim75$\,km\,s$^{-1}$. 

If we assume an inclination of $i\sim80^{\circ}$, the deprojected velocities would be a factor of $\sim \frac{1}{\cos{i}} \sim 5.8$ times larger assuming the orientation of the outflow is fixed perpendicular to the disk. This would mean the true velocity traced by the 6.2\,$\mu$m PAH feature would be $\gtrsim400$\,km\,s$^{-1}$; however, the orientation of the outflow material at such small scales close to the disk is not fully constrained. 

For both M82 and NGC 253, we find that the 1D velocity profiles along the outflow direction do not cross zero velocity at the kinematic center (labeled ``Original'' in Fig.~\ref{fig:VelProfiles}) unlike the disk velocities. For both targets, we found that the the kinematic center was $\sim5$pc different and so we have altered the x-axis position in Fig.~\ref{fig:VelProfiles} to reflect where the velocity profiles cross zero velocity (labeled ``Recentered'' in Fig.~\ref{fig:VelProfiles}).

\section{Discussion}
\label{sec:Discuss}

In both M82 and NGC 253, we observe clear velocity differences between different emission features in both the disk and outflow launching regions. In the disk of M82, we find that the H$_2$ S(1) line reaches lower velocities than the S(5) line. The ionized gas tracers are consistent with each other and show similar velocities to the PAHs and the H$_2$ S(5) rotational line. In NGC 253, a similar picture emerges, where the ionized lines show a different velocity to the molecular gas. Similar to M82, the PAHs (for NGC 253 only the 5.2~$\mu$m PAH shows a rotation signature) more closely trace the ionized gas rather than the molecular gas in their kinematics. 

Considering the high inclination of these galaxies, the line of sight velocity averages over different radii and so differences in the velocity curve may reflect overlapping contributions along the line of sight. It is possible that extinction plays a role, where different emission features are tracing different regions based on how obscured the line is, where the 
S(1) line may originate from regions with more dust as it is less affected by extinction than shorter wavelength lines. Alternatively, we might be tracking different physical conditions, where the hot molecular gas, as traced by the S(5) line, is tracing feedback regions where cavities in the gas are produced. In M82, there is a clear trend from the S(1) to S(3) to S(5) line, where the velocity increases, matching that of the ionized gas once the S(5) line is reached.

The consistency of the PAHs and ionized gas suggests we see PAH emission in the same regions, likely the ionized gas layer from PDRs while the cooler molecular gas is from more embedded regions, which may be impacted more by extinction and thus average to a different rotational velocity. 

A similar picture appears along the outflow launching direction, where the ionized gas shows higher velocities than the molecular gas ($\sim30$\,km\,s$^{-1}$ higher), consistent with theoretical predictions from simulations and analytical models \citep[][]{Schneider2020, Fielding2022, Richie2026}. Moreover, CO observations from \citealt{Nico2019} report lower molecular gas velocities ($\sim50-100$\,km\,s$^{-1}$) in the outflow of NGC 253 compared to the ionized gas velocities ($\sim200-400$\,km\,s$^{-1}$) as reported by \citealt{Cronin2025}. It is worth noting that both these works are probing the larger extended outflow rather than the launching region probed by our JWST spectroscopy.

\begin{figure}
\includegraphics[width=\columnwidth]{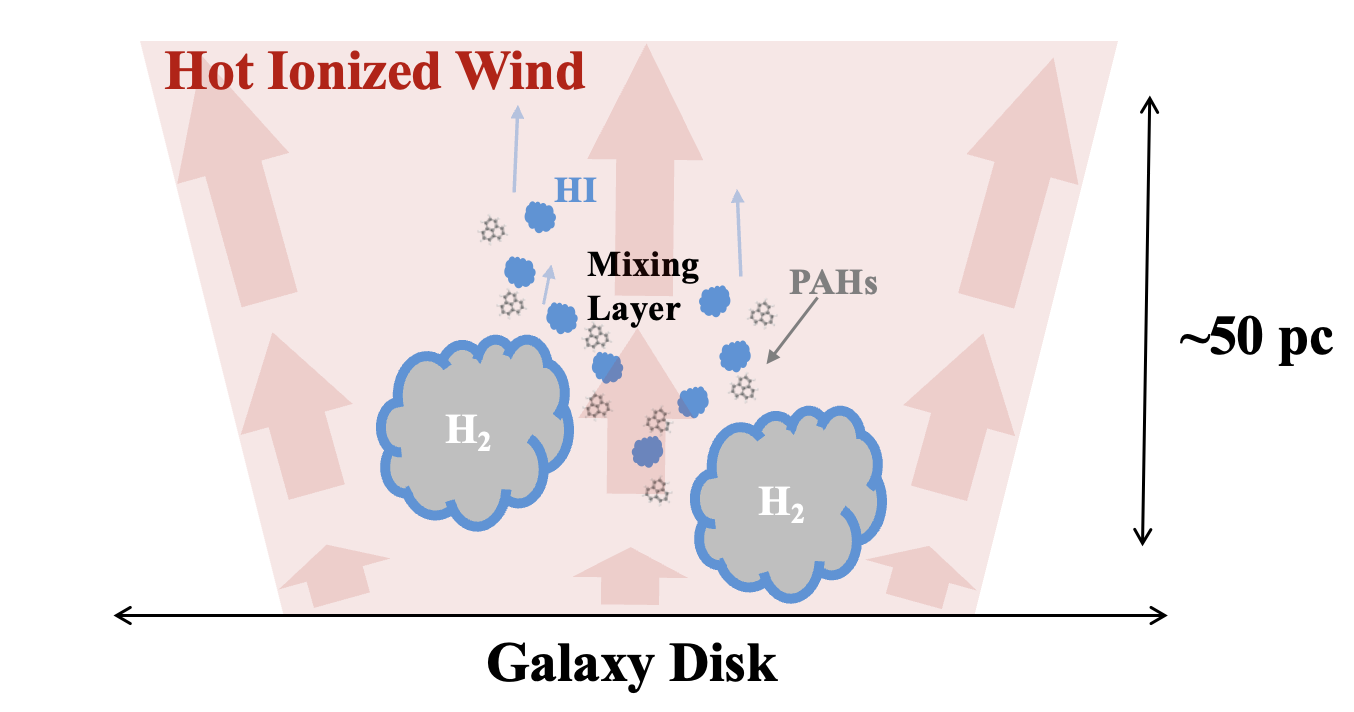}
    \caption{Outflow scenario consistent with our observations. Within the $\sim$50 pc region at the base of the outflow, molecular clouds are accelerated slowly while the outer layers of ionized gas and PAHs mix with the fast hot ionized wind, resulting in higher velocities of HI and PAHs compared to H$_2$. Additionally, the most mixed PAHs may be preferentially ionized, leading to higher velocities of the 6.2\,$\mu$m PAH feature.}
    \label{fig:Diagram} 
\end{figure}

This velocity difference is likely a product of a multiphase outflow, where dense molecular clouds are accelerated within a hot ionized wind. Due to the high inertia of the clouds, they cannot reach as high velocities within the $\sim30$ pc region where our observations probe. The emission from ionized gas originates from the outer layers of these clouds where they mix with the hot wind as they are accelerated \citep[e.g.][]{Westmoquette2009, Leroy2015, Gronke2018, Fielding2022}. We provide an illustration in Fig.~\ref{fig:Diagram}. Analytical models \citep[e.g.][]{Fielding2022}, show a similar picture, where the cold molecular clouds are slowly accelerated, with velocities of $\sim 50$ km~s$^{-1}$ within the $\lesssim300$ pc base of the outflow, while the ionized gas emission comes from the turbulent radiative mixing layer, reaching $100$s km~s$^{-1}$.

The rise and fall of the line of sight velocity as shown in the lower right panel of Fig.~\ref{fig:VelProfiles}, peaking at $\sim15$pc, either reflects the clouds decelerating or the direction of the velocity changes. After the cloud has accelerated it may disperse and is therefore decelerated by the interaction with the bulk material. At these small spatial scales close to the disk of the galaxy, we are possibly tracing the progenitors of the outflow, such as supernovae inflated bubbles, rather than the more ordered large scale outflow. As these expanding bubbles collide with one another, they will decelerate, explaining the slowdown beyond $\sim 15$pc.

Simulations from \citealt{Richie2026} predict that the majority of the dust mass is concentrated deep within the molecular clouds that are moving slowest, where PAHs are expected to get destroyed in the hottest phases of the outflow. Our results are not inconsistent with this picture, however, we find that the observed PAH velocity is $\sim30$\,km\,s$^{-1}$ higher than the H$_2$, where the 5.2\,$\mu$m $\&$ 11.3\,$\mu$m features are consistent with the ionized gas and the 6.2\,$\mu$m feature is significantly faster. This suggests that while the bulk of the dust mass may be within the molecular clouds, there is a fraction of PAHs that are in the mixing layer alongside the ionized gas (see Fig.~\ref{fig:Diagram}). Considering PAHs are predominantly excited by UV photons \citep[e.g.,][]{Li2002, Tielens2008} to emit, the PAHs locked inside the molecular clouds are not illuminated while those PAHs that are from the outskirts of the clouds are being entrained into the faster moving outflow. \citealt{Yoshida2019} used spectropolarimetry to measure the kinematics of dust grains in the wind of M82, and similarly they found a discrepancy between the fastest dust components and the molecular gas implying some kind of decoupling.

\citealt{Bolatto2024} reports a strong spatial correlation between Pa$\alpha$ and PAH emission in M82 in the larger scale outflow ($\sim$ kpc) while \citealt{Lopez2025} and \citealt{Fischer2025} also finds that PAHs are correlated with the ionized gas as traced by H$\alpha$, consistent with the picture shown in Fig.~\ref{fig:Diagram}. However, \citealt{Lopez2025} also finds that the PAH emission remains strongly correlated with CO throughout the wind of M82 up to 3.2~kpc, showing no deviations from galaxy disk CO-PAH relations. The CO-PAH correlation is not necessarily inconsistent with our results and may be a product of the flux of the PAH and CO depending primarily on the distance from the excitation source, in this case the galaxy disk. Indeed both \citealt{Lopez2025} and \citealt{Cronin2026} report that the flux of the PAHs and CO fall as $1/r^{2}$ from the galaxy disk (at least in the inner regions), although \citealt{Lopez2025} finds that the CO-PAH correlation holds after removing the dependence on distance. Moreover, as previously mentioned, the morphology of the PAHs matches that of the ionized gas, as traced by H$\alpha$ better than the CO \citep[see Fig.~4 of][]{Lopez2025}. A proper comparison of the CO velocities requires further work as the spatial scales probed are larger in the 1-0 and 2-1 transitions while the high resolution 3-2 data shows very little outflow contribution \citep[][]{Nico2019}. Future JWST/MIRI MRS observations would ideal to clarify the relationship between CO and the PAHs in the larger scale outflow.  

The 6.2\,$\mu$m PAH emission shows the highest velocity and acceleration in both galaxies, more-so than any of the ionized gas emission lines. This may suggest that we are tracing PAHs that are more mixed or fully mixed with the hot bulk wind than the 11.3\,$\mu$m and 5.2\,$\mu$m PAHs. Considering that the 6.2\,$\mu$m feature arises from ionized PAHs \citep[e.g.,][]{Draine2007, Rigopoulou2021}, this may suggest that we are seeing PAHs preferentially ionized after mixing into the wind that have been accelerated to higher velocities. Moreover, \citealt{Cronin2026} finds that the PAHs are predominantly ionized in the full wind of M82. It is worth noting, however, that PAHs are not expected to survive in the hot wind where thermal sputtering is efficient \citep[][]{Richie2026}, perhaps suggesting that PAHs are continuously supplied to the ionized phase from the shielded molecular clumps.

The outflows present in M82 and NGC 253, largely driven by supernovae \citep[e.g.][]{Veilleux2005, Heckman2017}, show a clear difference to AGN driven outflows. Unlike in AGN outflows, the 6.2~$\mu$m PAH feature is much more strongly detected, but also has the highest velocity of any of the emission features, as shown in Fig.~\ref{fig:VelProfiles}. Recent analysis of PAHs in AGN outflows show a weak 6.2\,$\mu$m PAH feature \citep[e.g.][]{Zhang2024, Garcia-Bernete2024b, Garcia-Bernete2026} suggesting more neutral PAHs. Furthermore, applying the same PCA tomography technique finds differences in the intrinsic profile rather than Doppler shifts \citep[][]{Donnan2026, Donnan2026c}. While there is ionization of the PAHs within starburst driven outflows, perhaps the harder radiation field of AGN outflows (where high ionization potential lines such as [\ion{Ne}{5}] are present), fully destroys the ionized PAHs rather than carrying them along with the outflowing gas.


\section{Conclusions}
In this work we have applied PCA tomography to measure the kinematics of PAHs in M82 and NGC 253, where we are able to produce high quality velocity maps of various PAH features as well as the molecular gas via H$_2$ rotational transitions, and ionized gas via [\ion{Ne}{2}] and Pf$\alpha$. Our key results are
\begin{itemize}

    \item We find that the kinematics of M82 are dominated by a rotating disk, where the PAHs are consistent with the ionized gas (as traced by [\ion{Ne}{2}]) and molecular gas as traced by the S(5) transition of H$_2$. The lower J transitions of H$_2$ appear to rotate more slowly, suggesting they trace different gas than the higher J H$_2$ lines.

    \item We find that the kinematics of NGC 253 show a mix of outflow and disk, with the 6.2\,$\mu$m and 11.3~$\mu$m PAHs dominated by the outflow while the molecular gas traces a rotating disk. The ionized gas shows both a clear outflow component and a disk component. It is worth stressing that the outflow scales are small ($\sim30$pc) compared to the $\sim$kpc extent, meaning we probe the launching within the starburst region.

    \item In NGC 253, we find that the outflow velocity of the ionized gas is higher than the warm molecular gas, with the PAHs more consistent with the ionized velocities. This is consistent with a picture where dense molecular clouds are launched and accelerated slowly within a faster hot bulk flow, where the PAH emission and ionized gas emission originate from a mixing layer between the cloud and wind, resulting in higher velocities.

    \item For both M82 and NGC 253, we find the 6.2\,$\mu$m PAH reaches higher velocities than the other PAH features and the ionized gas. This is perhaps due to preferential ionization of the PAHs that have mixed into the faster moving bulk flow of the wind.

\end{itemize}
This work demonstrates the value and power of kinematics to disentangle different phases of the ISM which can now be applied to PAH features. 

\newlength{\twocolwidth}
\setlength{\twocolwidth}{\columnwidth}

\begin{acknowledgments}
We thank the reviewer for their thorough and constructive review of the paper. FRD thanks Francisco Rodr\'iguez Montero and Dimitra Rigopoulou for the helpful discussion.
This work is based on observations made with the NASA/
ESA/CSA James Webb Space Telescope. We acknowledge
support from program JWST-GO-01701, which is provided by
NASA through a grant from the Space Telescope Science
Institute, operated by the Association of Universities for
Research in Astronomy, Inc., under NASA contract NAS
5-03127. The data were obtained from the Mikulski Archive
for Space Telescopes at the Space Telescope Science Institute,
operated by the Association of Universities for Research in
Astronomy, Inc., under NASA contract NAS 5-03127 for JWST
FRD and KS acknowledges funding support from grant JWST-GO-05279.002. 
E.A.C.M.  gratefully  acknowledges  funding  from the National  Science  Foundation  under  Award  Nos. 2206509, and CAREER 2339670.
R.S.K. acknowledges financial support from the ERC via Synergy Grant ``ECOGAL'' (project ID 855130) and from the German Excellence Strategy via the Heidelberg Cluster ``STRUCTURES'' (EXC 2181 - 390900948). In addition R.S.K. is grateful for funding from the German Ministry for Economy and Energy (BMWE) in project ``MAINN'' (funding ID 50OO2206), and from DFG and ANR for project ``STARCLUSTERS'' (funding ID KL 1358/22-1). V.V. acknowledges support from the Comité ESO Mixto 2024 and from the ANID BASAL project FB210003. R.H.-C. thanks the Max Planck Society for support under the Partner Group project "The Baryon Cycle in Galaxies" between the Max Planck for Extraterrestrial Physics and the Universidad de Concepción. R.H-C. also gratefully acknowledge financial support from ANID - MILENIO - NCN2024\_112 and ANID BASAL FB210003. L.A.B. acknowledges support from the Dutch Research Council (NWO) under grant VI.Veni.242.055 (\url{https://doi.org/10.61686/LAJVP77714}).

\facility{JWST}
\software{Astropy \citep{Astropy2013, Astropy2018, Astropy2022}, MatPlotLib \citep[][]{MatplotLib2020}, NumPy \citep[][]{Numpy2020}, SciPy \citep[][]{SciPy2020}, AplPy \citep[][]{aplpy2012, aplpy2019}}

\end{acknowledgments}

\newpage
\appendix

\section{Comparison to line fitting}
\label{sec:LineFits}

To verify that the PCA tomography technique is able to accurately determine kinematics, we check the velocity maps we infer for the emission lines with a more traditional approach of modeling the spectrum of each spaxel with a model Gaussian. In particular, we allow the central wavelength, amplitude and width of a Gaussian to vary as free parameters for each spaxel. The velocity of a given spaxel is inferred from the shift in the central wavelength.

\begin{figure}
	\includegraphics[width=\textwidth]{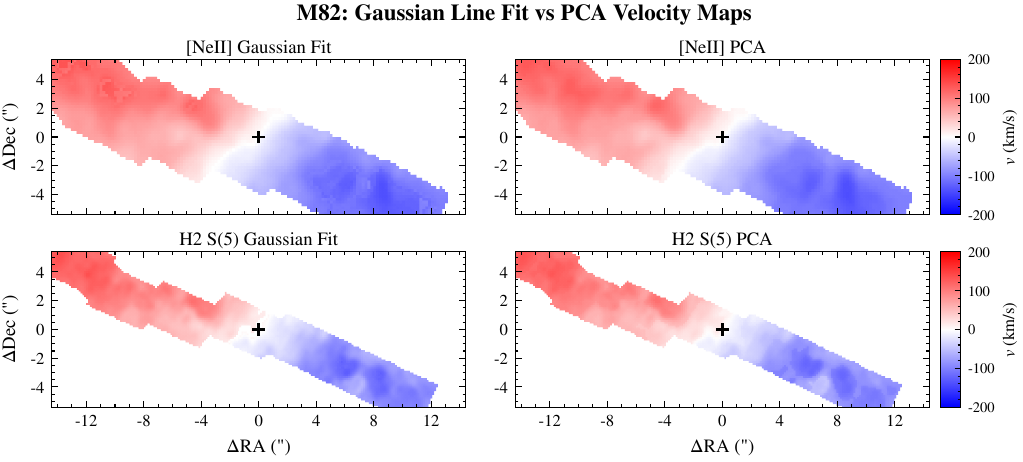}
    \caption{Comparison of the velocity maps inferred using the traditional method of fitting a Gaussian to every spaxel compared to the PCA technique. We show the [\ion{Ne}{2}] and H$_2$ S(5) lines for M82. This test demonstrates the robustness of PCA tomography to measure velocities. The median difference in velocity between the two maps is $|\Delta v|\lesssim3$km s$^{-1}$. }
    \label{fig:VelCompare} 
\end{figure}

Fig.~\ref{fig:VelCompare} shows this comparison for [\ion{Ne}{2}] and H$_2$ S(5), where we find that the velocity maps are indistinguishable demonstrating that PCA tomography is effective at producing velocity maps. We find that the median of the residuals between the Gaussian fitting and PCA technique are $|\Delta v|\lesssim3$km s$^{-1}$. 

Moreover we reproduce Fig.~\ref{fig:VelProfiles} from the line fitting (without the PAHs) and find the same velocity profiles where the H$_2$ S(1) line reaches a lower velocity than the S(5) line and the ionized gas lines.

\section{Variations in the Intrinsic Profile}
\label{sec:IntProfile2}
As presented in Section \ref{sec:IntProfile}, we can distinguish variations in the intrinsic profile from genuine kinematics by inspecting both the tomogram of the second principal component (to check if it appears similar to the known velocity maps) but more crucially by inspecting the eigenspectrum of the second principal component. 

\begin{figure}
	\includegraphics[width=\textwidth]{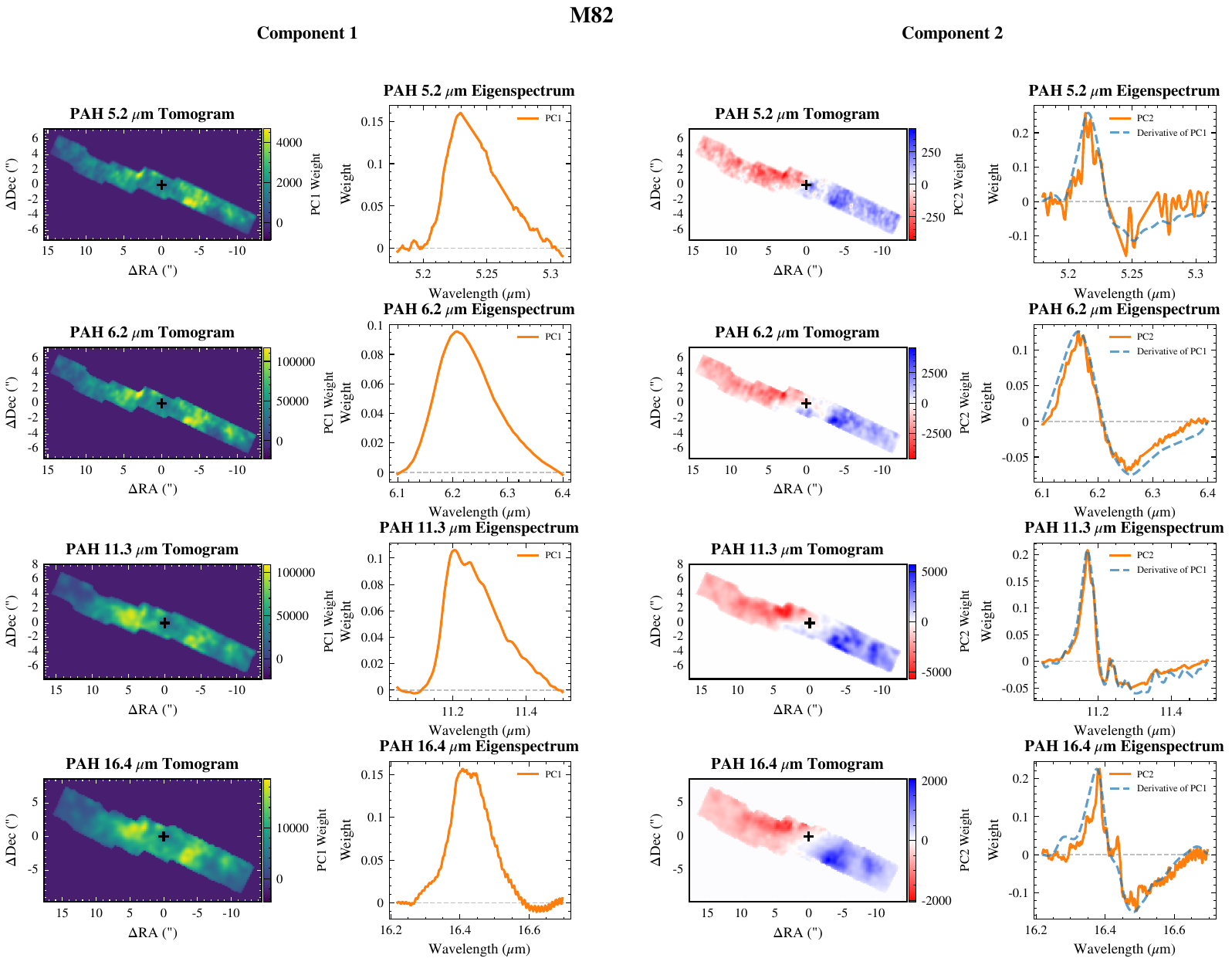}
    \caption{The first and second principal components of the PAH features for M82. The first two columns show the first component, with the left panels showing the tomograms while the right panels show the corresponding eigenspectrum (orange). The second component is shown in the final two columns, where the dashed blue line shows the derivative of the first principal component eigenspectrum, which should match the second principal component in the case of a Doppler shift. This plot demonstrates that velocity shifts dominate the profile changes rather than variations in the intrinsic profile.}
    \label{fig:2ndCompM82} 
\end{figure}

\begin{figure}
	\includegraphics[width=\textwidth]{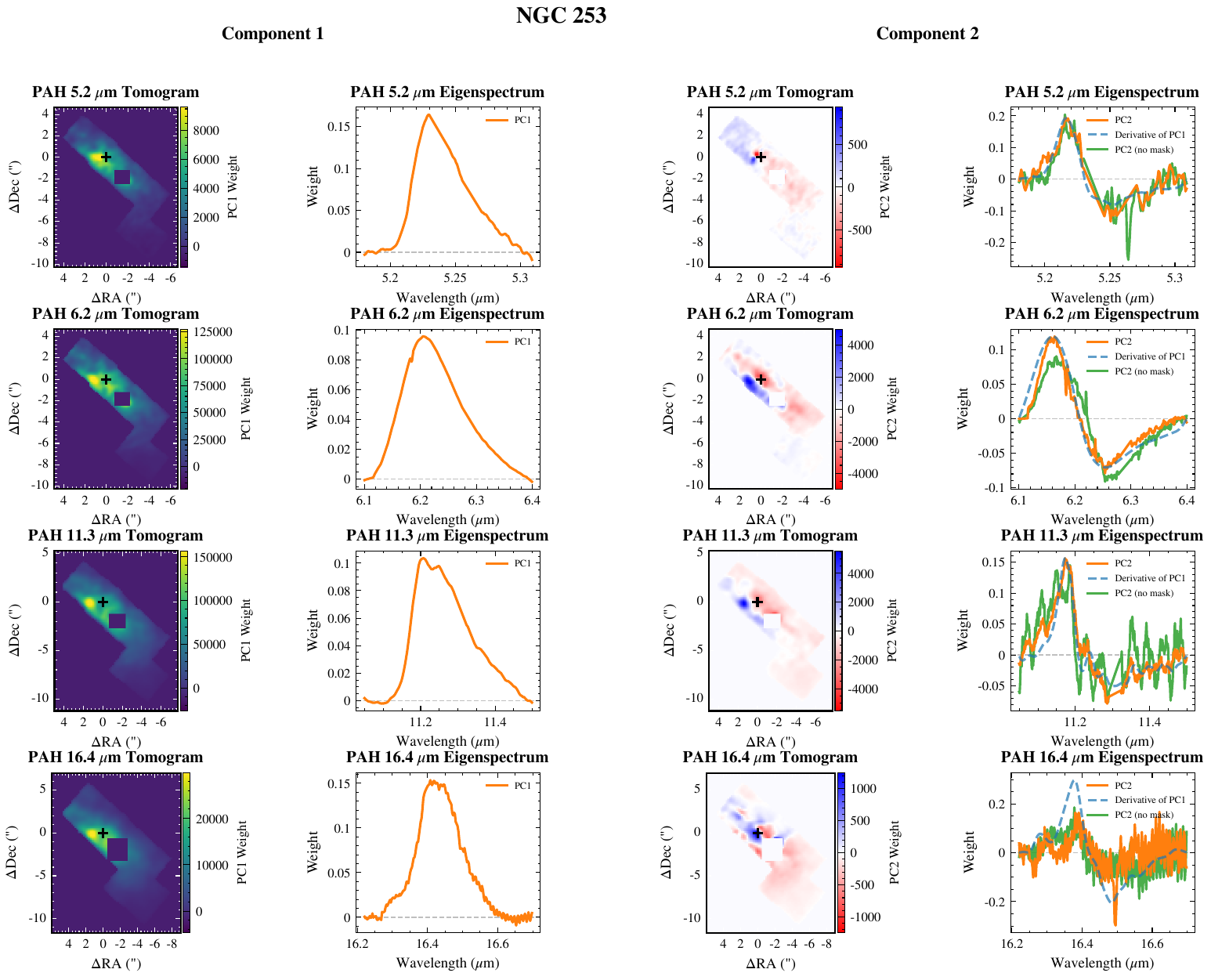}
    \caption{The same as Fig.~\ref{fig:2ndCompM82}. For NGC 253, we mask the cluster region and so we also show the second principal component where the cluster region is not masked (green), where we the 6.2\,$\mu$m feature shows differences in the intrinsic profile in this region. 
}
    \label{fig:2ndCompNGC253} 
\end{figure}

We show the first and second principal components of the 5.2\,$\mu$m, 6.2\,$\mu$m, 11.3\,$\mu$m and 17.0\,$\mu$m PAH features for M82 in Fig.~\ref{fig:2ndCompM82}, and for NGC 253 in Fig.~\ref{fig:2ndCompNGC253}. The first principal component traces the rest frame emission of the PAH feature. The second principal component is tracing kinematics, as the eigenspectrum matches the derivative of the first principal component. For NGC 253, we also show the second principal component if the cluster region was not masked in green. In this case we find that the second principal component does not match the derivative of the first, particularly for the 6.2\,$\mu$m PAH which suggests some variations in the intrinsic profile in that specific region. We also find that the 17.0\,$\mu$m PAH feature is too noisy for this target to produce a velocity measurement.

\begin{figure}
	\includegraphics[width=\textwidth]{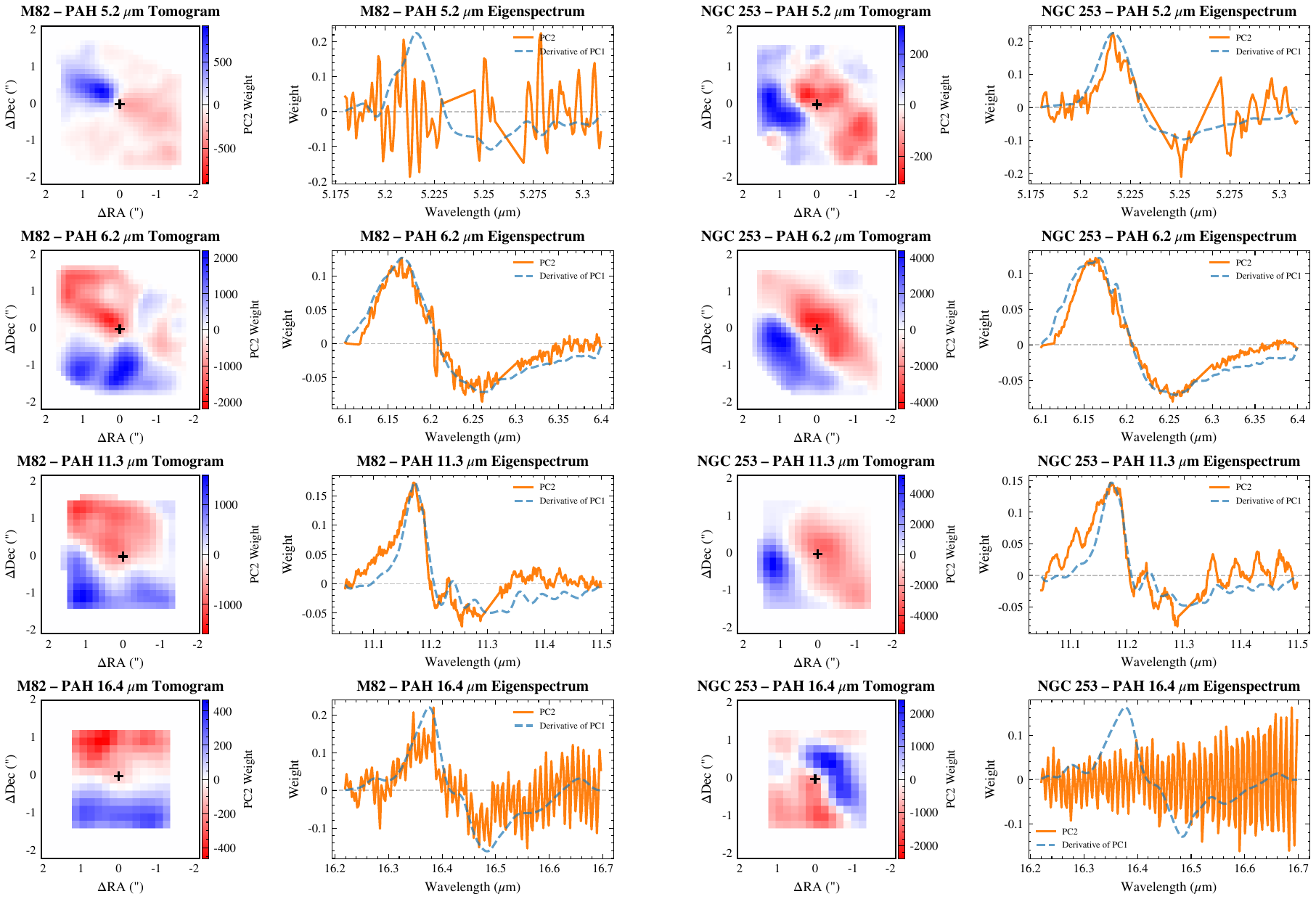}
    \caption{Same as Figures~\ref{fig:2ndCompM82} and \ref{fig:2ndCompNGC253} but for sub cubes of $2''\times2''$ regions around the nucleus to isolate the outflow regions. Note that the restricted field of view causes the 5.2\,$\mu$m PAH feature to fail for M82 as well as the 17.0\,$\mu$m PAH feature for NGC 253 - see Section \ref{sec:Mock} for discussion on how the FOV affects the sensitivity of PCA to detect kinematics of the PAHs. }
    \label{fig:2ndComp2} 
\end{figure}

As the principal component is determined from the full field of view of the data, we wanted to verify if isolating the outflow region also produces principal components consistent with a Doppler shift rather than differences in the intrinsic profile. To do this we formed new data cubes by taking a $2''\times2''$ region around the nucleus of each galaxy. The resulting second principal components are shown in Fig.~\ref{fig:2ndComp2}.

We find that the derivative of the first component still matches the second eigenspectrum when isolating the outflow regions, particularity for the 6.2\,$\mu$m feature. The 11.3\,$\mu$m PAH shows some differences for M82 compared to the full FOV which may suggest some variations in the intrinsic profile. We therefore further investigate whether these differences can cause the velocity differences we observe in our velocity maps. To do this we measure the difference in velocity when using the first two components of the PCA compared to using the derivative of the first component. To reconstruct the data cube to measure the velocity, we first project the derivative of the first eigenspectrum on the original data to calculate a tomogram using equation \ref{eqn:PCA}. 

\begin{figure}
	\includegraphics[width=\textwidth]{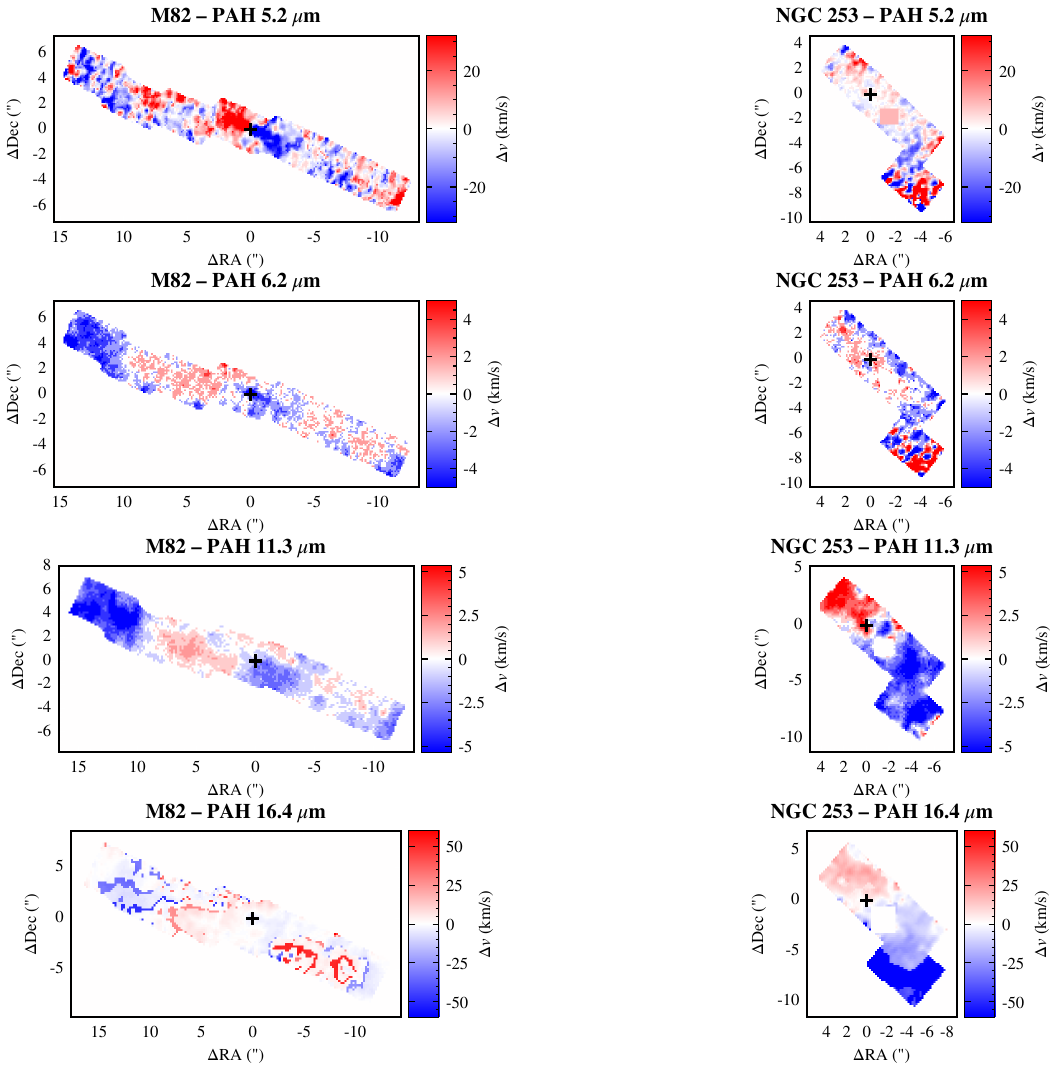}
    \caption{Velocity residual maps when taking the difference between the cube reconstructed using the first two principal components compared to using the derivative of the first principal component in place of the second principal component. This test shows that the 6.2\,$\mu$m and 11.3\,$\mu$m feature velocities are reliable to $\lesssim5$\,km\,s$^{-1}$.}
    \label{fig:VelDiff} 
\end{figure}

The difference in the velocity maps are shown in Fig.~\ref{fig:VelDiff}. These maps show where and how much the velocity may be incorrectly estimated to be due to differences in the intrinsic profile in M82 and NGC 253. We find that for the 6.2\,$\mu$m and 11.3\,$\mu$m PAH features that the magnitude of the velocity differences are $\Delta v \lesssim5$\,km\,s$^{-1}$, meaning the velocity differences we observe in Fig.~\ref{fig:Prof} are indeed real.

One final test is to apply the PCA to a data set where there are known variations in the intrinsic profile of the PAH features. The MIRI MRS data of the Orion bar from the PDRs4All Early Release Science program \citep[][]{Berne2022} is ideal, as the profile of all the PAH features has been shown to vary \citep[][]{Chown2023, Pasquini2023, VanDePutte2025} while the velocity is negligible.  

\begin{figure}
\centering                                           
	\includegraphics[width=10cm]{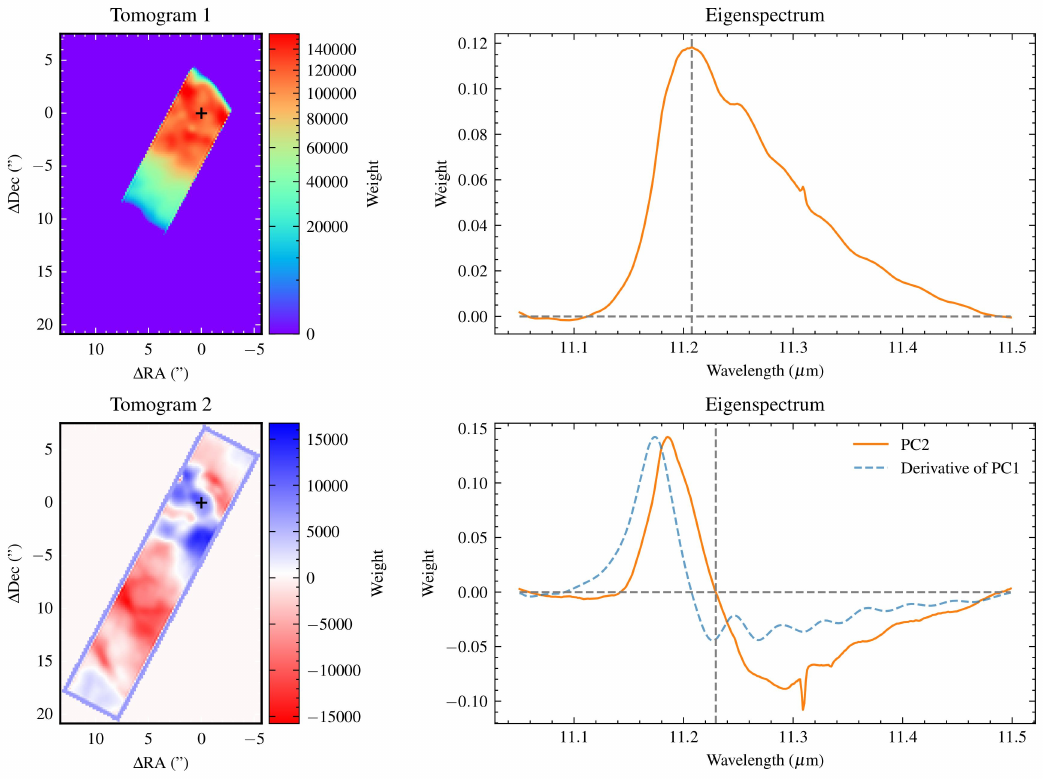}
    \caption{First two principal components of the 11.3\,$\mu$m PAH in the Orion bar - a region with no velocity but rather intrinsic variations in the emission profile. The cross marks the position of proto-planetary disk 203-504 as a reference point. The disagreement between the derivative of the first eigenspectra and the second shows that the shift in the intrinsic profile from Class A to Class B as defined by \citealt{Peeters2002} is different compared to a Doppler shift.}
    \label{fig:Orion} 
\end{figure}

We show the first two principal components for the 11.3\,$\mu$m PAH feature in Orion bar in Fig.~\ref{fig:Orion}, where the derivative of the first principal component is clearly different to the second principal component. The shape of the second eigenspectrum still describes a feature that is shifted in wavelength, however, the type of shift is different than a Doppler shift, reflecting a different intrinsic profile. Where the eigenspectrum is positive, the feature has more strength in the blue wings while the opposite is true where it is negative. The tomogram shows that the atomic PDR is consistent with a bluer feature while the region behind where the disassociation fronts are present show a redder feature. This is what was found by \citealt{Chown2023}, where the feature moves from Class A to Class B  \citep[][]{Peeters2002} along the bar from the Atomic PDR to the disassociation fronts, where the red wing of the 11.3\,$\mu$m PAH feature becomes more pronounced.

\section{Mock Data Tests}
\label{sec:Mock}
To determine the effectiveness of PCA tomography to detect the kinematics of PAH features, we performed some tests with mock data. One key variable we want to quantify is the effective spectral resolution that can be obtained for the PAH features using PCA tomography. How well the PCA can infer the kinematics of the PAHs will not only depend on the signal to noise of the data but also the number of spatial pixels (and thus the field of view), as the spatial exes contain the information used to decompose the data cube into principal components.

We generate a mock data set of MIRI Channel 2 which contains the 11.3\,$\mu$m PAH feature, using a PAH template (see Fig.~1 of \citealt{Donnan2026b}) generated from the average of star-forming regions from NGC 3256 and NGC 7469 \citep[][]{Donnan2024, Rigopoulou2024} from the GOALS ERS program (ERS 1328, P.I. L. Armus and A. Evans) extracted using \textsc{SPIRIT} \footnote{\url{https://github.com/FergusDonnan/SPIRIT}} \citep[][]{Donnan2024, Donnan2026_SPIRIT_zenodo}, a mid-IR fitting tool to model the PAH features and continuum. We generate a 2D Gaussian as the flux distribution with a radius of 10 pixels which corresponds to 1.7$''$. We then rotate and incline the 2D Gaussian by 45$^{\circ}$. We generate a velocity field using the disk model in \citealt{Donnan2026}, which has a position angle $\phi$ = 45$^{\circ}$, and inclination $i$ = 45$^{\circ}$, parameterized by a radial rotation curve with the form
\begin{equation}
\label{eqn:RotCurve}
    V(R) = V_{\textrm{max}}\tanh{\left(\frac{R}{R_{\textrm{turn}}}\right)},
\end{equation}
where the $V_{\textrm{max}}$ and $R_{\textrm{turn}}$ are free parameters describing the max rotation velocity and the radius where the curve flattens at $R>R_{\textrm{turn}}$. We fix $R_{\textrm{turn}}=1''$ for the purposes of this test. The velocity map, $V_{\rm obs}(x,y)$, is therefore
\begin{equation}
    V_{\rm obs}(x,y) = V(R)\,\sin i \,\cos\theta.
\end{equation}
The radius, $R$, is the deprojected radius which relates to the coordinates of each spaxel $(x,y)$ via
\[
\begin{aligned}
x' &= x\cos\phi + y\sin\phi, \\
y' &= -x\sin\phi + y\cos\phi, \\
R &= \sqrt{x'^2 + \left(\frac{y'}{\cos i}\right)^{2}}, \\
\cos\theta &= \frac{x'}{R}. \\
\end{aligned}
\]

With this parametrization of the velocity field, we can use the parameter $V_{\textrm{max}}$ as an indication of the velocity resolution but it is worth noting that the individual velocities of the spaxels can be lower than this. For our test we want to know, what is the minimum value of $V_{\textrm{max}}$ for a given signal to noise and field of view, can PCA tomography detect the kinematics of the 11.3\,$\mu$m PAH feature. 

We generate the mock data for a variety of signal to noise values, specifically, SNR = $[5,10,15,20,30,40]$ where the SNR is defined as the signal to noise of the peak flux of the 11.3\,$\mu$m PAH feature rather than the integrated flux. We also generate mock data for a variety of field of view (FOVs) where FOV = $[4, 6, 8, 10]$ arcseconds (diameter), as the PCA decomposition is more effective the more spatial pixels are available.

For each SNR and FOV, we generate 10 mock data cubes, with $V_{\textrm{max}} = [20, 40, 60, 80, 100, 120, 140, 160, 180, 200]$ km/s. We then run the PCA analysis for all 240 mock data cubes with the goal of determining what the minimum value of $V_{\textrm{max}}$ is for a given FOV and SNR. 

To define a successful detection of kinematics by the PCA decomposition, we first visually inspect the outputs of the test where we confirm that at low SNR and/or FOV, the PCA fails to recover the kinematics if $V_{\textrm{max}}$ is low. To quantify the success or failure of the PCA decomposition, we measure the sum of the square difference, $\chi^2$, of the eigenspectrum of the second component with respect to the derivative of the first component. If the PCA detects kinematics, these should match as discussed in Section \ref{sec:IntProfile} and Appendix \ref{sec:IntProfile2}. 

\begin{figure}
	\includegraphics[width=\textwidth]{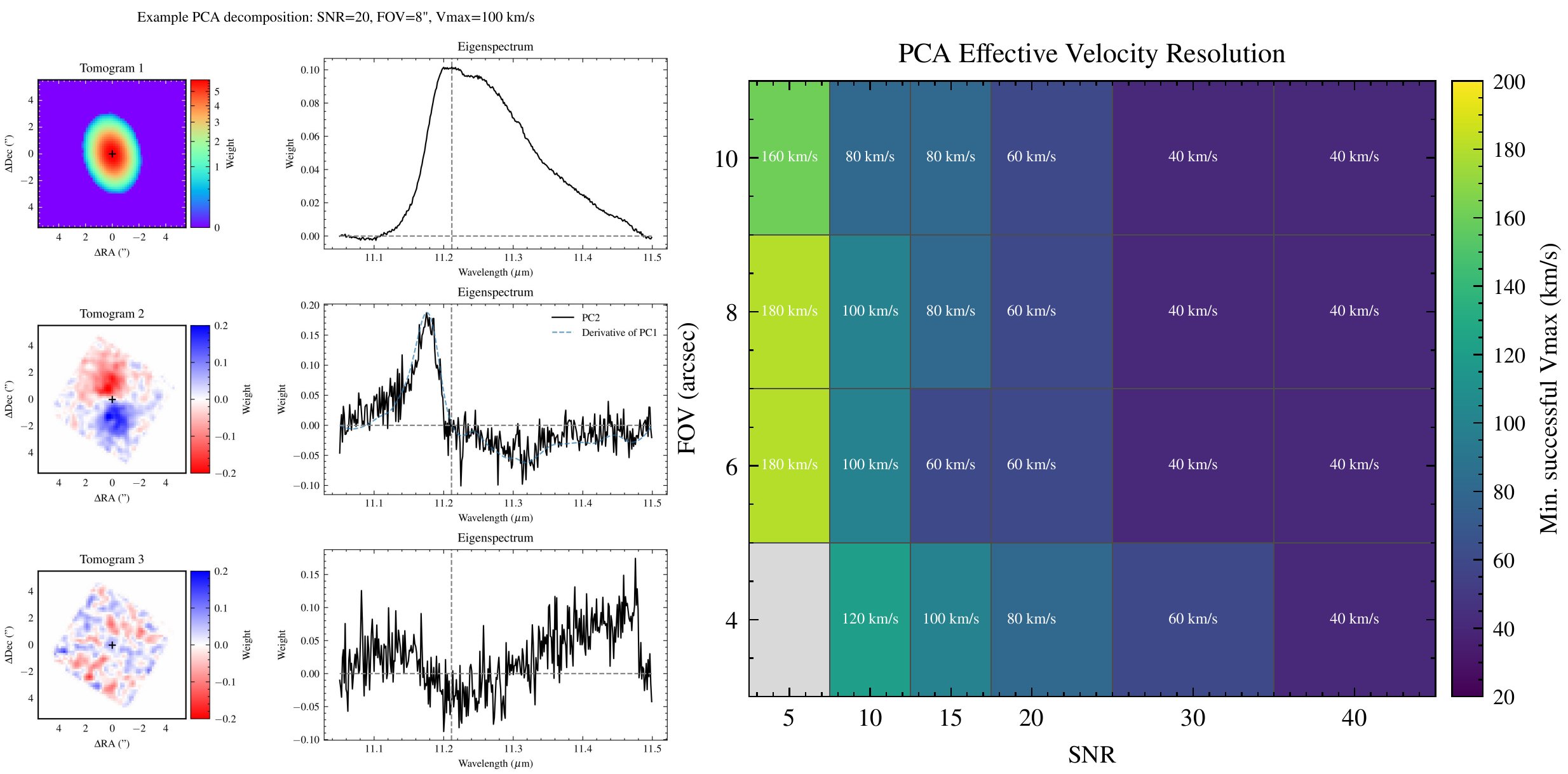}
    \caption{\textbf{Left:} Example PCA decomposition with the mock data, where each row shows each principal component with the tomogram on the left and the corresponding eigenspectrum on the right. The dashed blue line shows the derivative of the first eigenspectrum which matches the second eigenspectrum where a Doppler shift is detected. \textbf{Right:} The effective velocity resolution of the 11.3\,$\mu$m PAH feature as measured by PCA tomography of mock data of a given field of view and signal to noise. The gray panel in the bottom left means that no values of $V_{\textrm{max}}$ produced a successful kinematic detection for the lowest SNR and field of view for $V_{\textrm{max}}\le 200$ km/s. The M82 data has a median SNR$\sim155$ and a field of view of $\sim20''\times4''$ for reference. }
    \label{fig:VelRes} 
\end{figure}

We indeed observe that the $\chi^2$ decreases with increasing $V_{\textrm{max}}$ for each FOV and SNR and define a threshold value of $\chi^2 = 135$ for a successful detection of kinematics by the PCA decomposition. We show the minimum $V_{\textrm{max}}$ for each FOV and SNR in Fig.~\ref{fig:VelRes}. As expected we find the minimum $V_{\textrm{max}}$ to decrease with increasing FOV and SNR giving us a measure of the effective velocity resolution of the 11.3\,$\mu$m PAH. 

The M82 data has a FOV$\sim20''\times4''$ with a median signal to noise of SNR$\sim155$ meaning the quality of the data is more than sufficient to measure the kinematics of the PAH features with comparable accuracy to the emission lines, placing the effective velocity resolution at $\lesssim40$\,km\,s$^{-1}$. Note that this is the max velocity given a rotation curve parameterized by equation \ref{eqn:RotCurve} and so the difference in velocity from spaxel to spaxel can be lower than this value.

\bibliography{References}{}
\bibliographystyle{aasjournalv7}



\end{document}